\documentclass[prl,aps,twocolumn,showpacs,superscriptaddress,floatfix]{revtex4}
\usepackage{graphicx}
\usepackage[dvips]{color}
\begin{document}
\title{Supersolidity and phase diagram
of softcore bosons in a triangular lattice}
\author{Jing-Yu Gan}
\affiliation{Center for Advanced Study, Tsinghua University,
Beijing, 100084, China}
\author{Yu-Chuan Wen}
\affiliation{Institute of Theoretical Physics and
Interdisciplinary Center of Theoretical Studies, Chinese
Academy of Sciences, Beijing 100080, China}
\author{Yue Yu}
\affiliation{Institute of Theoretical Physics,  Chinese
Academy of Sciences, Beijing 100080, China}

\begin{abstract}
We study the softcore extended Bose Hubbard model in a
two-dimensional triangular lattice by using the quantum
Monte Carlo methods. The ground state phase diagram of the
system exhibits a very fruitful structure. Except the Mott
insulating state, four kinds of solid states with respect to
the commensurate filling factors $\rho=1/3,2/3$ and $\rho=1$
are identified. Two of them (CDW II and CDW III) are newly
predicted. In incommensurate fillings, superfluid,
spuersolid as well as phase separation states are detected .
As in the case for the hardcore bosons, a supersolid phase
exists in $1/3<\rho<2/3$ while it is unstable towards the
phase separation in $\rho<1/3$. However, this instability is
refrained in $2/3<\rho<1$ due to the softening of the bosons
and then a supersolid phase survives.

\end{abstract}


\maketitle

\noindent {\it Introduction} A quantum matter state with
non-zero order parameters characterizing both solid and
superfluid phase, so-called supersolid (SS), was proposed by
Penrose and Onsager fifty years ago\cite{penrose}. In 1970s,
it was speculated that the large zero-point quantum
fluctuation in solid $^4$He may induce an
off-diagonal-long-range-order \cite{andreev}. The recent
observation of non-classical rotational inertia (NCRI) in
porous Vycor or in bulk solid $^4$He reported by Kim and
Chan\cite{kim} have greatly revived the interest in studying
this new matter state although it is still controversial
whether the mechanism suggested by Refs. \cite{andreev}
works in explaining the NCRI \cite{con}.

\begin{figure}[b]

\vspace{-.6cm}

\includegraphics*[width=7cm]{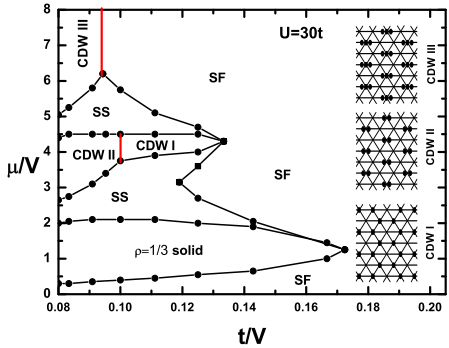}
\caption{(Color online) The ground state phase diagram for
the extended Bose Hubbard model in the two-dimensional
triangular lattice($U=30t$). SF is standing for the
superfluid phase; MI for the Mott insulator phase. The
$\rho=1/3$ solid means one of three sets of sublattices is
occupied with one atom per site. In the inset, the three
different solid states are illustrated. In CDW I at
$\rho=2/3$, two sets of sublattices are occupied with one
atom per site. In CDW II at $\rho=2/3$, one set of
sublattices is occupied with two atoms per site; In CDW III
at $\rho=1$, a single set of sublattices is occupied with
three atoms per site.} \label{phase1}
\end{figure}

Comparing to the complication of the $^4$He, the dilute cold
atom gas loaded in an optical lattice \cite{op} is a more
promising avenue to success the SS, especially when the
long-range interacting atoms were already cooled to a
Bose-Einstein condensate \cite{di}. These systems may be
well described by the Bose-Hubbard model \cite{bh} and its
extended version. The simple Bose-Hubbard system does not
allow a SS phase for the lacking of the interaction-based
long range correlation. It was also checked that there is no
a SS phase in the hardcore Bose-Hubbard model with a nearest
neighbor repulsion in two-dimensional square lattice due to
the solid-superfluid phase separation(PS), while a strip
supersolid presents if the next nearest neighbor interaction
is switched on \cite{batouni}. The lattice frustration may
stabilize a supersolid against the PS. This has been shown
in the hardcore Bose-Hubbard model in a triangular lattice
around the half filling \cite{hardtri1,hardtri2, hardtri3,
hardtri4} whereas the supersolid order is not favored in
more complicated frustrated lattices such as Kagom\`{e}
lattice \cite{kagome}. The next nearest neighbor interaction
in triangular lattice can also lead to  a strip supersolid
in triangular lattice \cite{striptri}.

Besides the next nearest neighbor interactions and lattice
frustration, softening of the on-site interaction may also
stabilize the supersolid phase because more than one atoms
per site are allowed, which has been confirmed either for a
two-dimensional square lattice \cite{softcore1} or
one-dimensional lattice \cite{softcore2}.

\begin{figure}[t,b]
\vspace{-.5cm}
\includegraphics*[width=9.0cm,height=3.5cm,angle=0]{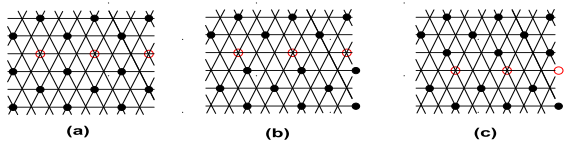}
\caption{(Color online) The $\rho=1/3$ solid doped by holes.
(a) Placing the doped holes into one line costs no
additional potential energy; (b) shifting the lower part of
the lattice introduces a domain wall without any potential
energy cost; (c) the doped holes (atoms) can hop freely
across the domain wall. } \label{fig:pd1}
\end{figure}

\begin{figure}[t,b]
\includegraphics*[width=7cm,angle=0]{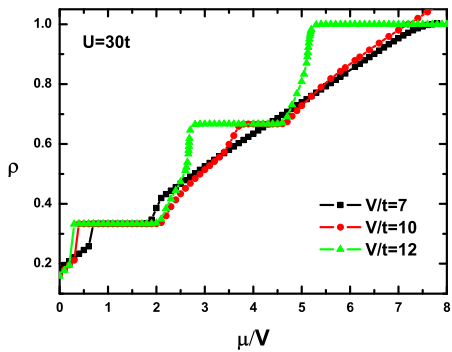}
\caption{(Color online) The average atom density as a
function of the chemical potential for $V=7t, 10t$ and
$12t$, respectively($U=30t$). The system size is $L=12$.}
\label{fig:pd2}
\end{figure}

\begin{figure*}[ht!!!]
\vspace{-.5cm}
\includegraphics*[width=15cm,height=7cm]{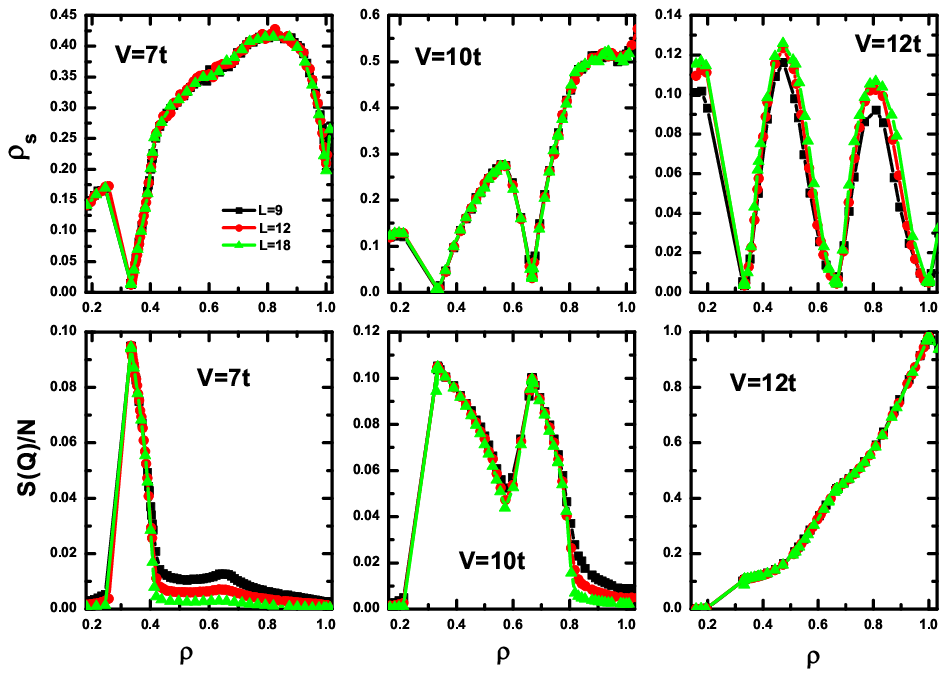}
\caption{ (Color online) Static structure factor $S(Q)$ and
superfluid density ($\rho_s$)as functions of atom density
$\rho$ for $V=7t, 10t$ and $12t$, respectively($U=30t$).
$Q=(4\pi/3,0)$ for triangular lattice.} \label{fig:pd3}
\end{figure*}

\noindent{\it Model and explanation to results} As we have
seen, the physics of the supersolid phase is determined by
the interplay among the next nearest neighbor interaction,
the frustration of lattice as well as the softening of the
on-site interaction. In this paper, we would like to check
the cooperating effect of latter two. The model we study is
the extended Bose-Hubbard model in a two-dimensional
triangular lattice, which reads
\begin{eqnarray}
H & = &-t\sum_{\langle ij\rangle}(a_{i}^{\dagger
}a_{j}+a_{j}^{\dagger }a_{i})
+\frac{U}{2}\sum_{i}n_{i}(n_{i}-1) \nonumber\\
&& +V\sum_{\langle ij\rangle}n_{i}n_{j} - \mu\sum_i
n_i,\label{hami}
\end{eqnarray}
where $a_i^{\dagger }(a_i)$ is the creation(annihilation)
operator of bosonic atom at site $i$;
$n_i=a_{i}^{\dagger}a_i$ is the occupation number; $\mu$ is
the chemical potential. $\langle ij\rangle$ runs over
nearest neighbors. $U$ and $V$ represent the on-site and
nearest neighbor repulsive interactions, respectively. Our
central result is presented in Fig.\ref{phase1}, the ground
state phase diagram. The quantum Monte Carlo(QMC) simulation
with the stochastic series expansion(SSE)
\cite{sandvik1,sandvik2} was applied.
 We use
$t$ as the energy unit and focus our calculations on
$U=30t$.

This phase diagram can be qualitatively explained by
strong-coupling arguments\cite{softcore1, hardtri1}. When
$U\rightarrow \infty$, the Hamiltonian (\ref{hami}) reduces
to the hardcore one. The phase diagram is symmetric about
$\rho=1/2$, i.e., $\mu/V=3$, because of the particle-hole
symmetry \cite{hardtri1}. The CDW II and CDW III as well as
the upmost SS are ruled out. The phase transition from the
solid to the SF is the first order one for the solid-SF PS
appears at the phase boundary. Finite $U$ breaks the
particle-hole symmetry and results in other two solid phases
and one more SS phase. The phase diagram with $t=0$ is quite
easily understandable. A meaningful chemical potential is
$\mu>-U/2$ otherwise $\rho \leq 0$. Raising $\mu$, there
would be several kinds of solid phases with a common
commensurate wave vector ${\mathbf Q}=(4\pi/3,0)$. The first
one is the $\rho=1/3$ solid in which one set of sublattices
is occupied with single atom per site. At $\rho=2/3$, there
are two kinds of solids depending on the ratio $V/U$. For
small values of nearest neighbor repulsion $V$($V/U<1/3$),
two sets of sublattices are occupied with single atom per
site (CDW I), while for $V/U>1/3$, one set of sublattices is
occupied with two atoms per site (CDW II). For $\rho=1$, the
solid phases involved are the MI and the other one. When
$V/U>1/3$, one set of sublattices is occupied with three
atoms per site (CDW III). If $V/U<1/3$, the MI, i.e., single
atom per site in the whole lattice, is the most stable. For
$\rho>1$, these commensurate solid phases are repeated in
filling factors with an integer added.

Based on this strong coupling limit understanding, we can
describe the phase diagram we obtained. Comparing to the
hardcore boson phase diagram in ref. \cite{hardtri1}, the
$\rho=1/3$ solid lobe change slightly with only the up-bound
lowered a little due to quantum fluctuations. When holes are
introduced into the $\rho=1/3$ crystal, the domain-wall
proliferation mechanism, i.e., the additional holes hop
freely across the domain wall, raises the kinetic energy
linearly in t (see Fig. \ref{fig:pd1}) and then excludes the
possibility of a SS phase~\cite{hardtri1,softcore1}. This
leads to a PS between $\rho=1/3$ solid and a uniform
superfluid with $\rho<1/3$~\cite{zhang}.

When atoms are introduced into the $\rho=1/3$ crystal, the
domain-wall proliferation mechanism does not work because
formation of a domain wall would cost extra potential
energy. Thus, a SS phase is there right above this solid
lobe, similar to that in the hardcore boson case. However,
it completely separates the $\rho=1/3$ and $\rho=2/3$ solid
lobes. Furthermore, the latter shrinks much and splits into
CDW II and CDW I in the $V^{-1}$ direction due to the ratio
$V/U$ is lowered.

Doping atoms into the $\rho=2/3$ solid, the system behaviors
differently in terms of the ratio between $V$ and $U$.
Following an analysis in Ref. \cite{softcore1} for the
formation of a SS phase above the half-filling in a square
lattice, we argue that a stable SS exists for $\rho>2/3$.
For $V/U<1/3$, CDW I is doped. Since the lattice sites are
either single occupied or empty, the energy cost for an
additional atom fills in empty or single occupied site is
$E^{I}_0=6V-\mu$ or $E^{I}_1=U+3V-\mu$, respectively. The
energy of a single atom delocalized between these two cases
is $E^I=E^I_0+\Delta^I-\sqrt{\Delta^{I2}+(6t)^2}$ with
$\Delta^I=(U-3V)/2$. Similarly, for $V/U>1/3$, i.e., doping
in CDW II, an additional atom fill either in an empty or
double-occupied site. The corresponding energy cost is
either $E^{II}_0=2U-\mu$ or $E^{II}_2=6V-\mu$, respectively.
The kinetic energy of a single particle delocalized between
these two cases is
$E^{II}-E^{II}_0=\Delta^{II}-\sqrt{\Delta^{II 2+}(6t)^2}$,
with $\Delta^{II}=(3V-U)/2$. In both cases, when $\Delta\sim
t$, the kinetic energy $E-E_0$ is proportional to $t$.
Therefore, the domain wall formation was blocked and the
system is stabilized against the PS \cite{softcore1}. As a
result, these doped atoms form a SF on the top of the
$\rho=2/3$ solid background, which is thought as a SS phase.
Notice that the above analysis is not applied to
$U\to\infty$ because multiple occupation is not allowed. The
domain wall forms if doping atoms into the $\rho=2/3$ solid
and then a PS happens \cite{hardtri1}.

For $\rho=1$, the ground state is a solid with one set of
sublattices occupied with three atoms per site (CDW III).
Doping holes into this state, these additional holes move
under the solid background with the effective hopping
$t^*\sim t^2/(7V-2U)$. The kinetic energy gain remains
quadratic in $t$ and small. Therefore, the SS state with
$\rho>2/3$ is stable. However, our numerical result does not
fully confirm this argument as we shall see later.

\noindent{\it Quantum Monte Carlo simulations} We now
portray our QMC simulations. In order to characterize
different phases, we study the static structure factor
$S(Q)$ with $Q=(4\pi/3,0)$ and superfluid density $\rho_s$,
\begin{eqnarray}
S(Q) =  \frac{1}{N}\sum_{ij}e^{iQ\cdot(r_i - r_j)} \langle
n_i n_j\rangle, ~~~ \rho_s  = \frac{\langle
W^2\rangle}{4\beta t}
\end{eqnarray}
where $W$ is the winding number fluctuation of the bosonic
would lines \cite{pollock}; $\beta=2L$ is the inverse of
fictitious temperature and $N=L\times L$ is the lattice
size. In our calculations, we use $L=9,12$ and 18.  A solid
phase is characterized by a non-zero $S(Q)$ and $\rho_s=0$
and a SF phase by $S(Q)=0$ and $\rho_s\neq 0$; a SS phase is
depicted by non-vanishing both of $S(Q)$ and $\rho_s$.
Solids are categorized by their $S(Q)$ values.

Fig. \ref{fig:pd2} shows the atom density $\rho$ varying as
the chemical potential $\mu$ for $V=7t, 10t$ and  $12t$,
respectively($U=30t$). It is seen that for each $V$, there
is a jump from $\rho<1/3$ to $\rho=1/3$ which indicates a
first order phase transition. In the grand canonical
ensemble, a jump in $\rho$ is a token of  a PS region. For a
small $V$, say $V=7t$, only $\rho=1/3$ solid plateau is
observed, i.e., there is no $\rho=2/3$ plateau. This is
different from that in the hardcore boson phase diagram  for
in that case there are both the $\rho=1/3$ and $\rho=2/3$
solid lobes symmetrically. Obviously, this is because the
particle-hole symmetry is broken by the finite $U$. As $V$
is raised, say $V=10t$ and $12t$, the $\rho=2/3$ solid
plateau appears and is gradually widened. Such an
observation of the $\rho=2/3$
 solid plateau indicates the
shrinking of the $\rho=2/3$ solid lobe as shown in Fig.
 \ref{phase1}. In the regions $1/3<\rho<2/3$ and $2/3<\rho<1$,
both of them include a SS phase. For $V=12t$, another solid
plateau at $\rho=1$ emerges at larger $\mu$, while this
plateau is not observed for $V=7t$ and $10t$.

In Fig. \ref{fig:pd3}, we show $S(Q)$ and $\rho_s$ vary as
functions of $\rho$. At $\rho=1/3$, for all $V$ we considered, the
superfluid density $\rho_s\rightarrow 0$ while $S(Q)/N\sim 1/9$
which is an exact magnitude of the static structure factor of the
$\rho=1/3$ solid state. For $\rho>1/3$, the results are severely
dependent on the magnitude of $V$. For a small, say $V=7t$,  as
$\mu$ increases, we observe, in turn, a $\rho=1/3$ solid plateau,
a SS phase in the region $1/3<\rho<0.4$ and a SF after $\rho>0.4$.
No CDW III solid state is observed at $\rho=1$. For an
intermediate value of $V (=10t)$, both $S(Q)$ and $\rho_s$ are
finite and non-monotonous as $\rho$ raises. A SS phase expands to
the whole region from $\rho=1/3$ to $\rho=2/3$. At $\rho=2/3$,
$\rho_s\rightarrow 0$ while $S(Q)/N\sim 0.1$ which is the
magnitude of the static structure factor of CDW I. With further
increasing of $\rho$, there is another SS in between
$2/3<\rho<0.8$. After $\rho=0.8$, $S(Q)\rightarrow 0$, i.e., the
system is in a pure SF phase. Again, no CDW III exists at
$\rho=1$. For large $V$, say $V=12t$,  $\rho_s\rightarrow 0 $ at
$\rho=2/3$ while $S(Q)/N\sim 0.45$. This is CDW II solid state. As
$\rho$ increases further from $\rho=2/3$, $S(Q)$ raises
monotonously and both $\rho_s$ and $S(Q)$ in nonzero till
$\rho=1$. Thus, in $2/3<\rho<1$, the ground state is a SS phase.
At $\rho=1$, the ground state is another kind of solid state with
$\rho_s\rightarrow 0$ and $S(Q)/N\sim 1$, i.e., CDW III.

\begin{figure}[t,b]
\vspace{-.5cm}
\includegraphics*[width=7.5cm,angle=0]{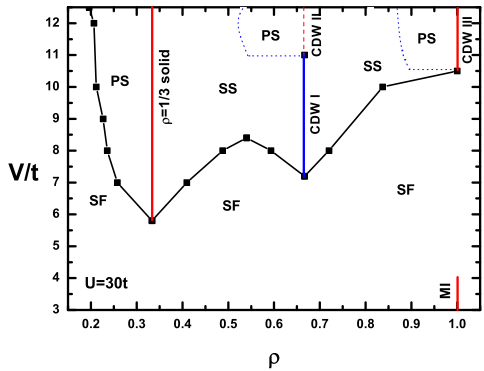}
\caption{(Color online) The ground state phase diagram for
two-dimensional extended Bose Hubbard model in the
triangular lattice in the $V-\rho$ plane at $U=30t$.}
\label{fig:pd4}
\end{figure}

\noindent{\it Phase diagrams and conclusions} We scan $V$
from $3t$ to $12t$ with a wider step. This sketches the
ground state phase diagram of the system . Fig. \ref{phase1}
and Fig. \ref{fig:pd4} are its projections to the
$\mu-$$V^{-1}$ and $V-\rho$ planes, respectively. The PS
phase is explicitly shown in Fig. \ref{fig:pd4} and its
width grows as $V$ is strengthened. On the left hand sides
of the CDW II and CDW III in Fig. \ref{fig:pd4}, we use dot
lines labelling two possible PS regions. These PS states are
neither supported by our early analysis nor confirmed by our
numerical data. However, our data could not exclude these
PS. What we see is although there seems no discontinuity in
$\rho$ of $\mu$, e.g., for $V=12t$ in Fig. \ref{fig:pd2},
the change of $\rho$ may be very sharp and for some specific
magnitudes of $V$ it is near a right-angle. We squint
towards there are no these PS  states but simulations with
larger lattice size and lower fictitious temperature are
required and confirmations with other numerical methods,
e.g., a canonical ensemble calculation, are necessary. No
such an ambiguity for the hardcore bosons because there are
no CDW II and CDW III states \cite{hardtri1}.

At $\rho=1$, for quite small values of $V$, say $V=3t$ and
$4t$, the MI phase( $\rho_s\rightarrow 0$ and
$S(Q)\rightarrow 0$) exists. There is a SF region in between
the MI and CDW III solid states and this region is
suppressed as $U$ becomes strong. Comparing with the
hardcore boson phase diagram depicted in Ref.
\cite{hardtri1}, we see that the softcore phase diagram is
much more fruitful. And on the other hand, it is more
practical to be verified by experiments because $U$ is
always finite. Comparing with the softcore bosons in a
square lattice \cite{softcore1}, the frustration of a
triangular lattice opens a wider window to explore the SS
state. The CDW III solid state at integer filling as the
counterpart of the MI state is another new feature in the
present model and is worth to be experimentally measured.
Mermin-Wagner theorem obstructs the various phases described
in the present work to be detected in any finite
temperature. However, it may become possible if we trap the
atoms by an additional shallow trapping potential as the
Bose-Eistein condensate is observed in a trapped
two-dimensional cold Bose atom gas.

This work was supported in part by Chinese National Natural
Science Foundation.

\end{document}